\documentclass{kapproc} 
\usepackage{graphicx}

\normallatexbib

\begin{document}

\articletitle{Continuum Field Model\\ of Defect-Induced
Heterogeneities\\in a Strained Thin Layer}

\author{Mikhail Belogolovskii, Gennadij Akimov, Yurii Komysa}
\affil{Donetsk Physical and Technical Institute\\
Donetsk 83114, Ukraine}
\email{bel@kinetic.ac.donetsk.ua}

\author{Paul Seidel}
\affil{Institut f\"{u}r Festk\"{o}rperphysik, Friedrich-Schiller
Universit\"{a}t\\ Jena D-07743, Germany}
\email{seidel@ifk.uni-jena.de}

\begin{abstract}
We investigate the effect of external stresses on structural and
mechanical properties of a strained damaged thin layer by
developing a continuum phase-field mesoscale model based on the introduction
of an order parameter field, the defect concentration, coupled
with a displacement field. We find that even in the case of an
initially uniform distribution of point defects external stresses 
drive the nucleation of local regions with higher concentration
of vacancies or self-interstitials than their average value over
the film. The effect can explain our experimental findings relating
generation of highly heterogeneous regions in cobalt disilicide film
fabricated in self-aligned processing on a silicon surface as well as 
improvement of fracture toughness in a tetragonal zirconia
ceramics with oxygen vacancies.
\end{abstract}


\section*{Introduction}
Material properties of thin solid layers are often controlled by
complex microstructure formed during their fabrication. Let us
assume a film with a uniform distribution of point defects that
is uniaxially strained. If lattice deformations are small (in
experiments discussed below they are limited to a few tenths of a
percent), according to naive expectations, the distribution of the
defects should remain the same in a strained state, i.e.,
homogeneous. The purpose of this paper is to prove that, in
general, it is not true and even comparatively small stresses may
result in any significant modification of a displacement field 
as well as the microstructure within the sample 
that can cause different mechanical and chemical changes on a
macroscopic scale. We address 
this issue by means of a novel
phase-field approach for a strained damaged layer based on the
introduction of two continuum fields relating the defect
concentration and atom displacements. The following section 
gives an outline of the theoretical approach and in 
remainder of the paper a simplified one-dimensional quasi-static model is
studied numerically. The data obtained are discussed with respect 
to two our experiments dealing with strained thin layers of
cobalt disilicide and zirconia ceramics that exhibit an 'unusual' sensitivity
of their material properties to (apparently weak) disorder. 

\section{The Model}
Phase-field methodology has become increasingly recognized as a useful 
tool for simulating mechanical failure due to external stresses and
understanding the role of the underlying physical processes (see the 
papers~\cite{Bhate,Aranson,Karma,Eastgate} and referencies therein). 
Advantages of the continuum approaches starting from basic theoretical 
assumptions have been clearly demonstrated: first, they avoid any abrupt
intefaces, facilitating numerical convergence and, second, they are
dealing with more realistic equations of motion for the material than 
alternative sharp interface theories. The diffuse interface models consider
amorphous solids and are based on the introduction of two fields: 
that of a local order parameter describing fluctuations in the mass
density and a standard displacement field ${\bf u}({\bf r})$ of mass 
points measured from their original positions. The latter one 
representing strain in the material is usually treated within the 
conventional linear elasticity theory whereas the former field
is supposed to obey a double-well Ginzburg-Landau potential with two minima
representing vacuum and perfect solid states (it turns out to be hard to
realize physical limitations of such an approximation that
covers the 
whole range of the order parameter between two limiting cases). 
In contrast to the works used two-field fracture models,
we focus here on small fluctuations of both parameters when
approximations made become more evident and physically clear.
We show that in apropriate circumstances (if some scale in the material
is comparatively small) the combination of weak disorder
and strain may change the system properties dramatically. 

In this section we present the basic ideas of the approach similar in
philosophy to, but very different in detail from, the works cited above. 
Local disorder will be characterized by the relative concentration $c({\bf r})$
of point defects in the material, positive for
vacancies and negative for self-interstitials;  $|c({\bf r})|<1$. 
We start from a non-strained 
solid with a homogeneous distribution of defects $c({\bf r})=c_0$ 
produced during its fabrication process when the system
passed through a sequence of metastabe states. If thermal fluctuations
are not relevant, it usually remains in a given configuration as long as
the state corresponds to a free energy minimum. The latter condition
means that any local deviation from $c_0$ will only increase the energy 
and for small
variations of $c({\bf r})$, we may limit ourselves with a quadratic term
in the free energy density of a damage solid, 
$k(c({\bf r})-c_0)^2/2$, where $k$ is a phenomenological parameter.
Following Ginzburg-Landau ideas on phase transitions, we introduce 
a gradient term $g({\nabla}c({\bf r}))^2/2$, energetically suppressing 
spatial fluctuations
in the order parameter. Another limiting case is a perfect crystal in a 
deformed state whose local elastic free energy density is assigned by the 
strain  tensor   
$u_{ik}=\partial{u_i}/\partial{x_k}+\partial{u_k}/\partial{x_i}$. 
Within the linear elasticity theory it is of a quadratic form 
$\lambda _{iklm}u_{ik}u_{lm}/2$ with parameters 
$\lambda _{iklm}$ that for a homogeneous, isotropic 
material are described by two Lam\'e constants~\cite{Landau} 
whereas for a system
undergoing a first order transition the energy
should be taken as a Landau polynomial
of strain tensor components.

What is less trivial is the coupling between 
two fields $c(\bf{r})$ and $\bf{u}(\bf{r})$. 
To give an insight about its form, we shall discuss a crystal containing a 
mesoscopic dilation center at a point ${\bf r}_0$
formed by self-interstitials. Its presence is regarded for a solid 
as an external origin of stresses causing displacements of nearest atoms
from their equilibrium positions. Then the stress tensor may be approximated 
with a simple formula $\sigma_{ik}=K\Omega _{ik}\delta ({\bf r}-{\bf r}_0)$, 
where $K$ is the bulk modulus, 
a linear combination of two Lam\'e constants. 
As $u_{ll}=\sigma_{ll}/3K$, the total change of the solid volume $\delta V$, 
that can be roughly estimated as the number of interstitials 
in the dilatation center multiplied by 
an atomic volume, is equal to $\delta V=\int u_{ll}\rm{d}V$~\cite{Landau}. 
Hence, the trace of the tensor $\Omega _{ik}$ is $\Omega _{ll}=3\delta V$. 
For a single dilatation center the interaction energy 
is equal to $-K\Omega _{ik}u _{ik}(\bf{r}_0)$ that can be generalized in the 
common case of a continuous defect distribution as $c({\bf r})\alpha _{ik}u_{ik}({\bf r})$, 
where $\alpha _{ik}$ is a tensor with components whose values can be very roughly estimated as the bulk modulus $K$. The second effect is related with changes of 
elastic characteristics $\lambda _{iklm}$ for host-host atom interactions
due to their displacements in the vicinity of defect: 
$\delta \lambda_{iklm}=\Omega^{\star}\Lambda _{iklm}\delta({\bf r}-{\bf r}_0)$, 
where $\Omega^{\star}$ is the value of the order of $\delta V$, 
parameters $\Lambda _{iklm}$ are positive for 
a dilatation center (host-host atom interactions become stronger 
in the nearest  neighborhood of the defect) and negative for a pore.
For a continuous distribution of local defects we may approximate 
$\delta \lambda_{iklm}$ as $-c({\bf r})\lambda _{iklm}$. 
An additional argument for such a substitution is that 
in the case of $c({\bf r})=1$ the elastic energy is vanishing.   
Phase-field equations describing a quasi-static behavior of the system (that is an 
aim of the paper) can be obtained by minimizing its total free energy. 

\section{Numerical Simulations and Discussion}
To show how our model is working and what can happen with a damage 
but homogeneous elastic solid under strain, we limit ourselves to a one-dimensional 
approximation. It means that we shall assume the presence of a wide and thin 
layer of a finite length 2$L$. Collecting all contributions to the free energy 
together and rescaling $x/L \longrightarrow x$, 
$g/(\lambda L^2) \longrightarrow g$,  $k/\lambda \longrightarrow k$, 
$\alpha /\lambda \longrightarrow \alpha$,  
we get the following expression for a total free energy of the solid:  
\begin{eqnarray}
\label{eq:F} 
F=L\lambda \int_{-1}^1dx\lbrack\frac{g}{2}(\frac{dc(x)}{dx})^2+
\frac{k}{2}(c(x)-c_0)^2+
\nonumber \\
+\alpha c(x)\frac{du(x)}{dx}+
\frac{1}{2}(1-c(x))(\frac{du(x)}{dx})^2\rbrack.
\end{eqnarray}        
In a quasi-static case spatial distributions of the defect concentration $c(x)$ and atom displacements $u(x)$ should minimize the free energy, i.e., 
$\delta F/\delta c(x)=0$ and $\delta F/\delta u(x)=0$. It yields
the following set of two differential equations:
\begin{eqnarray}
\label{eq:dF} 
g\frac{d^2c(x)}{dx^2}=k(c(x)-c_0)+\alpha \frac{du(x)}{dx}-\frac{1}{2}
(\frac{du(x)}{dx})^2;
\nonumber\\
\frac{d^2u(x)}{dx^2}=\frac{1}{1-c(x)}\frac{dc(x)}{dx}\lbrack\frac{du(x)}{dx}
-\alpha \rbrack.
\end{eqnarray}
The equations thereby incorporate the physics of the defect subsystem 
and the macroscopic behavior of a strained solid that follows from the 
conventional theory of elasticity. Because we are dealing with a 
finite-size problem, it is very important to formulate correct 
boundary conditions. First, we suppose that displacements at the 
boundaries are known: $u(\mp L)=\pm \delta L$ (it is positive when 
the sample is squeezed and negative for stretching efforts). 
Second, because of the symmetry of the problem the function $c(x)$ 
should be symmetrical on $x$. And, last, we assume a conserved 
order parameter, i.e., the initial number of defects will not be changed  
in a strained material: 
\begin{equation}
\label{eq:c} 
\int_{-1}^1c(x)dx=2c_0.
\end{equation}
Such a condition may be applied to a solid layer if defects (vacancies 
and self-interstitials) are already redistributed within it but the 
process of their annihilation at boundaries has not started yet. 

In Fig. 1a we show both fields $c(x)$ and $u(x)$ calculated for 
a positive value of $c_0$ and certain other reasonable parameters.
It follows that the concentration of vacancies at the 
center of the layer is larger than at its ends.
\begin{figure}[ht]
\centering
\includegraphics[scale=0.8]{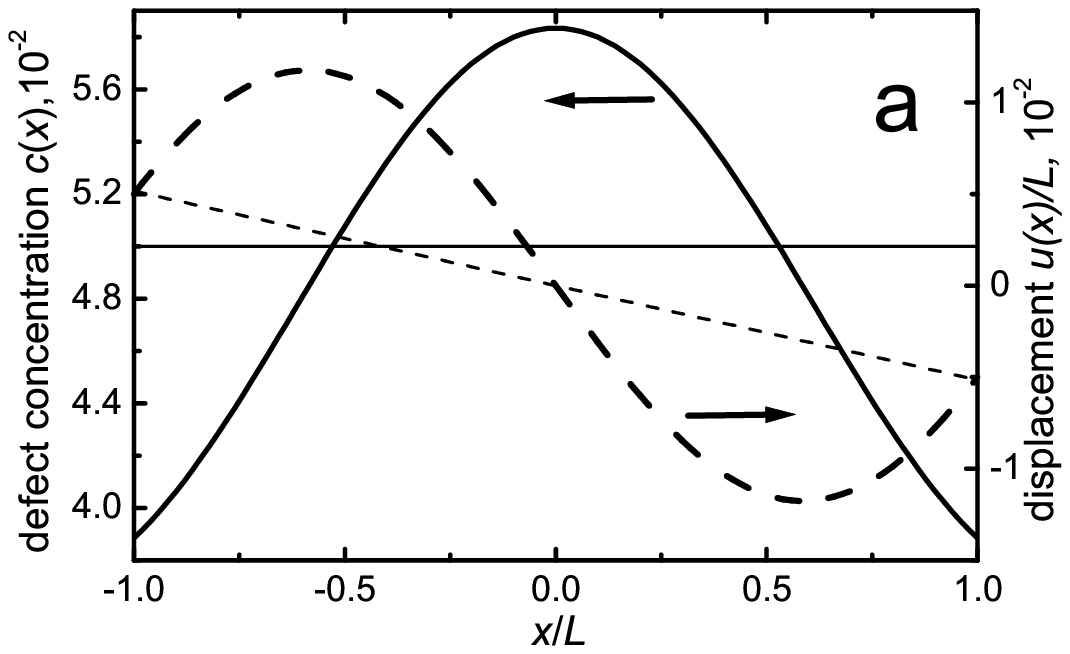}
\\\vspace{0.1in}
\includegraphics[scale=0.8]{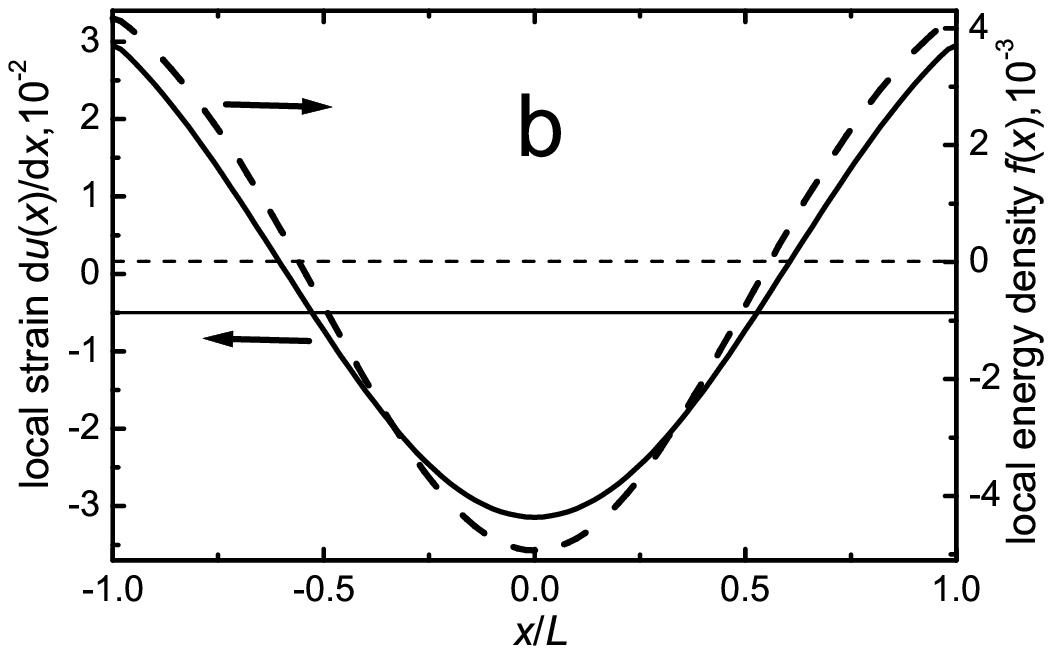}
\caption{(a) Profiles of the defect distribution $c(x)$ (solid line) and 
the displacement field $u(x)$ (dashed line). (b) The strain $du(x)/dx$ 
(solid line) and the local free energy density $f(x)$ (dashed line);
$\delta L/L$=0.005; $c_0$=0.05; $g$=1.0; $k$=3.0, and $\alpha $=3.0. 
Corresponding characteristics for a homogeneous film without any 
defects are shown by relating thin (solid and dashed) lines.}   
\end{figure}
The displacemenr field is also inhomogeneous that 
is more evident from Fig. 1b, where the stress $du(x)/dx$ (which characterizes relative
changes of the length along the sample) has maximal negative values 
in the middle part of the sample. Because the mass
of the film is fixed, large squeezing at the center is compensated by the
stretch at ends, where local relative changes of the length are positive (see Fig. 1b). 
The parameters were chosen to be similar to those for experiments described below.

A strong enhancement of the mechanical response 
induced in a material by a small perturbation at its boundaries
can imply 'unusual' phenomena  
if some scale in the system is comparatively small. 
To show where and how our model can be useful, we refer to
two our experiments with a combination
of weak disorder and small stresses. We start with a solid phase
reaction that is thermally initiated on a substrate covered by other
reagent and results in formation of the layer of 
a new compound. The reaction rate follows the Arrhenius law 
$\mathrm{exp}(-W/k_{\rm B}T)$ with an activation energy $W$, the Boltzmann constant
$k_{\rm B}$ and the temperature $T$, usually $W\gg k_{\rm B}T$. 
It should be emphasized that
even in the case of tiny variations $\delta W$ of the energy barrier to be overcome, 
their effect can be observable on a macroscopic scale if $\delta W$ is comparable 
with the temperature coefficient $k_{\rm B}T$. In this case  
significant reaction rate modifications (of several times) may take place 
in different parts of the heterostructure. We believe that it can be one of 
the main sources of the heterogeneity of CoSi$_2$, a practically important
material that is usually obtained in self-aligned silicide processing~\cite{Muraka}. 
The layer of cobalt disilicide is formed by depositing a Co layer onto a Si substrate
and heating the combination until the two elements react to form
Co$_2$Si, then CoSi, and finally the desired phase CoSi$_2$. The
cobalt disilicide films obtained by such a way are usually rough, contain 
a high density of pinholes and even break into islands~\cite{Belousov}. It raises a
question of either the heterogeneity is a purely technical problem, or
there is something more fundamental behind it. In our opinion, the mechanical 
failure is a sequence of substantial thermal stresses in a defect Co layer 
due to an essential mismatch of thermal expansion coefficients between 
silicon and cobalt. It leads to the overall relative Co film contraction of
nearly 0.5\% at 700$^o$C, the highest processing temperature, that, 
according to Fig.~1, drives nucleation of regions with
higher concentration of defects. 
Simultaneously, an inhomogeneous displacement field is
generated that results in local changes of the free energy (see Fig.~1b). 
In principal,
they are very small comparing with the reaction activation energy 
but (as our estimates show) are of the same order of magnitude as the temperature 
scale $k_{\rm B}T$. And it just may lead to the formation of highly 
heterogeneous CoSi$_2$ coatings during the silicidation 
reaction. The ways to avoid it are to reduce external stresses and/or
to get the cobalt film as clean as it is posssible. We have performed 
first experiments which show that the layer-by-layer procedure 
strongly reducing thermal stresses allows formation of CoSi$_2$ layers of
a better quality than those produced by a standard silicidation technique. 

Another experimental support for the model has 
come from our experiments for 4-mol\% yttria-stabilized
tetragonal zirconia polycrystals with intrinsic oxygen
vacancies. The nearly 4\% volume
expansion during the tetragonal t-ZrO$_2$ to monoclinic m-ZrO$_2$
phase change~\cite{Nishiyama} triggered in the crack vicinity 
is conventionally regarded as the main mechanism leading to strengthening in 
zirconias containing tetragonal grains. 
Our three-point bend measurements for stripes with 
a significant amount of the vacancies~\cite{Akimov} have revealed  
fracture toughness (resistance
to crack propagation) two times higher than the characteristic for
a usual tetragonal zirconia without defects. 
Because the transformation toughening is 
a process dominated by the volume increase, the
effect observed in damaged samples~\cite{Akimov} 
subjected to an external stretching force may be
interpreted in terms of an enhanced driving force for the
tetragonal to monoclinic inversion. Based on the concepts 
elaborated above, the latter mechanism 
could be explained as a generation of m-ZrO$_2$ precipitates
inside tetragonal grains. Really, calculations for a negative sign  
of $\delta L$ (similar to those of Fig.~1 where $\delta L$ 
is positive) show that in the case of expanded 
samples the local lengthening of a damaged film    
is greater in the center than at the ends. Moreover, if to imitate 
the proximity to a phase transition by taking the local elastic free
energy density of the system in a form of a dilational strain
Landau polynomial, the effect becomes more pronounced.
It means that already at apparently small strains the presence of defects
would favor the generation of regions with a significantly larger
atomic volume. We believe that just their presence initiates 
the tetragonal $\longrightarrow $ monoclinic transformation
and thus improves the mechanical characteristics.

In conclusion, within the framework of a continuum field model we have shown
that defects may induce a spatially inhomogeneous strain state in a
stressed film. In the light of present results, our experimental
findings for damaged layers of cobalt disilicide~\cite{Belousov} and 
4-mol\% yttria-stabilized zirconia~\cite{Akimov} are explained qualitatively.  
We do hope that our approach can be applicable to some other 
fields of thin film metallurgy and mechanics.

\begin{acknowledgments}
The authors wish to thank I.~Belousov and A.~Grib 
for valuable discussions relating experiments with
cobalt disilicide buffer layers and the German BMBF for 
its support within the joint German-Ukrainian WTZ-Project
UKR 01/051.
\end{acknowledgments}


\begin{chapthebibliography}{1}

\bibitem{Bhate} Bhate, D.~N., Kumar, A. and Bower, A.~F. (2000). Diffuse 
interface model for electromigration and stress voiding.
{\em J. Appl. Phys.}, 87:1712--1721.

\bibitem{Aranson} Aranson, I.~S., Katalsky, V.~A. and Vinokur, V.~M. (2000). Continuum field description of crack propagation.
{\em Phys. Rev. Lett.}, 85:118--121.

\bibitem{Karma} Karma, A., Kessler, D.~A. and Levine H. (2001). 
Phase-field model of mode III dynamic fracture.
{\em Phys. Rev. Lett.}, 87:045501-1--045501-4.

\bibitem{Eastgate} Eastgate, L.~O., Sethna, J.~P., Rauscher, M., 
Cretegny, T., Chen, C.-S. and Myers, C.~R. (2002). Fracture in
mode I using a conserved phase-field model.
{\em Phys. Rev. E}, 65:036117-1--036117-10.

\bibitem{Landau} Landau, L.~D. and Lifshitz, E.~M. (1964). 
{\it Theory of Elasticity}. Oxford: Pergamon Press. 

\bibitem{Muraka} Muraka, S. (2000). Silicidation, 
in {\it Handbook of Semiconductor Manufacturing Technology},
Y.Nishi and R.Doering, eds. New York: Marcel Dekker.

\bibitem{Belousov} Belousov, I., Rudenko, E., Linzen, S. and 
Seidel, P. (1997). Local nucleation and lateral crystallization 
of the silicide phases in the CoSi$_2$ buffer layer of
YBa$_2$Cu$_3$O$_{7-x}$/CoSi$_2$/Si structure.
{\em Microelectronic Engineering}, 37/38:581--587;
Belousov, I., Grib, A., Linzen, S. and Seidel, P. (2002).
Cobalt silicide formation inside surface defects of a silicon
substrate. {\em Nucl. Instr. and Meth. in Phys. Res. B}, 
186:61--65.

\bibitem{Nishiyama} Nishiyama, Z. (1978). 
{\it Martensitic Transformations}. 
New York: Academic Press. 

\bibitem{Akimov} Akimov, G.~Ya, Marinin, G.~A. and Kameneva, V.~Yu. 
(2003). Evolution of the phase content and physical-mechanical 
properties of the ceramics ZrO$_2$+4 mol.\% Y$_2$O$_3$. 
{\em Fiz. Tverd. Tela (St. Petersburg)} [{\em Phys. Solid State}], 
in press.
 
\end{chapthebibliography}
\end{document}